\documentstyle[12pt,epsf,twoside,fleqn,espcrc1]{article}

\newcommand{\be}{\begin{eqnarray}}
\newcommand{\ee}{\end{eqnarray}}

\newcommand{\AmS}{{\protect\the\textfont2
  A\kern-.1667em\lower.5ex\hbox{M}\kern-.125emS}}

\title{Understanding Bottom Production\thanks{This work was supported
in part by the Director, Office of Energy Research, Division of Nuclear Physics
of the Office of High Energy and Nuclear Physics of the U. S.
Department of Energy under Contract No. DE-AC03-76SF0098.}}

\author{N. Kidonakis\address{Department of Physics and Astronomy, 
University of Rochester,
Rochester, NY 
USA}, E. Laenen\address{NIKHEF Theory Group, 
1009 DB Amsterdam, The Netherlands}, 
S. Moch\address{Institut f\"{u}r Theoretische Teilchenphysik, Universit\"{a}t
Karlsruhe, 
Germany} and R. Vogt\address{Lawrence 
Berkeley National Laboratory, Berkeley, CA 
USA}$^,$\address{Physics Department, 
University of California, Davis, CA 
USA}}

\begin{document}

\maketitle

\begin{abstract}
We describe calculations of $b \overline b$
production to next-to-next-to-leading order (NNLO) and next-to-next-to-leading
logarithm (NNLL) near threshold in $pp$ interactions.  Our calculations are
in good agreement with the $b \overline b$ total cross section measured by 
HERA-B.
%We compare the NNLO-NNLL results
%to the existing $b \overline b$ total cross section data 
%and discuss the implications for the extrapolation to
%heavy ion collider energies.  
\end{abstract}\\

Factorization assumes it is possible to separate QCD cross sections into
universal, nonperturbative parton densities and a perturbatively calculable
hard scattering function, the partonic cross section.  However, some remnants
of long-distance dynamics in the hard scattering function can dominate
corrections at higher orders near production threshold.  These Sudakov
corrections have the form of distributions singular at partonic threshold.
Threshold resummation techniques organize the singular distributions to all
orders, extending the reach of QCD into near threshold production.  The
singular functions organized by resummmation are the plus distributions,
$[\ln^l x/x]_+$, where $x$ denotes the `distance' from partonic threshold.

The first attempts to resum the heavy quark cross section were at leading
logarithm (LL) and exploited the fact that to LL, the Sudakov corrections to
the heavy quark cross section were identical to those obtained for Drell-Yan
production \cite{LSvN,KSbot,SVchm}.  This early resummation calculation, like
some of the later results that followed, used an empirical cutoff to keep the
strong coupling constant from blowing up.
Resummation beyond LL cannot make use of this universality because the color
structure of each partonic process must be treated separately \cite{KS}.
%simple structure
%because the $gg$ contribution to $Q \overline Q$ production has a more 
%complex structure.  
The NLL $Q \overline Q$ terms were first resummed in
Ref.~\cite{KSV} for a simplified case 
%where the scattering angle between the
%$Q$ and $\overline Q$ was held fixed.  The entire angular range was 
and later fully
solved for the $q \overline q$ channel \cite{KV}.   The exponents in
the $gg$ channel were calculationally unwieldly, requiring large cutoffs
even for $t \overline t$ production where the resummation should work best.
However, one advantage of the resummed cross section is that when it is
expanded in powers of $\alpha_s$, it provides estimates of unknown
finite-order corrections without resorting to a cutoff or other prescriptions.
We have calculated the double-differential heavy
quark hadroproduction cross sections up to next-to-next-to-leading order
(NNLO), ${\cal O}(\alpha_s^4)$, and next-to-next-to-leading logarithm (NNLL),
{\it i.e.}\ keeping powers of the singular functions as low as $l = 2i - 1$ 
at order ${\cal O}(\alpha_s^{i+3})$ where $i=0,1, \ldots$ \cite{KLMV,nick}.
Since resummation is based on expansion of the LO cross section, we only
discuss $Q \overline Q$ production in the $ij = q \overline q$ and $gg$
channels since $qg$ scattering first appears at NLO.

In our calculations, the distance from partonic threshold in the singular
functions depends on how the cross section is calculated, either by integrating
away the momentum of the unobserved heavy quark or antiquark and determining
the one-particle inclusive (1PI) cross section for the detected quark, or by
treating the $Q$ and $\overline Q$ as a pair in the integration, pair invariant
mass (PIM) kinematics.   In 1PI kinematics,
\begin{eqnarray}
\label{eq:3}
p(P_1) + p(P_2) \longrightarrow Q(p_1) + X(p_X)\, , 
\end{eqnarray}
where $X$ denotes any
hadronic final state containing the heavy antiquark and
$Q(p_1)$ is the identified heavy quark.  
The reaction in Eq.~(\ref{eq:3}) is dominated 
by the partonic reaction
\begin{eqnarray}
\label{eq:7}
i(k_1) + j(k_2) &\longrightarrow& Q(p_1) +
X[\overline Q](p_2')\, . 
\end{eqnarray}
At LO or if $X[\overline Q](p_2') \equiv \overline Q(\overline p_2)$, 
the reaction is
at partonic threshold with $\overline Q$ momentum $\overline p_2$.  
At threshold the
heavy quarks are not necessarily produced at rest but with
equal and opposite momentum.
The partonic Mandelstam invariants are
\begin{eqnarray}
  \label{eq:9}
s =(k_1+k_2)^2 \, ,\quad t_1 = (k_2-p_1)^2 -m^2 \, ,\quad
u = (k_1-p_1)^2 -m^2 \, ,\quad s_4 = s + t_1 + u_1 \, 
\end{eqnarray}
where the last, $s_4 = (p_2')^2 - m^2$, is the inelasticity of the partonic
reaction.  At threshold, $s_4=0$.  Thus the distance from threshold in 1PI
kinematics is $x = s_4/m^2$.  In 1PI kinematics, the cross sections are
functions of $t_1$ and $u_1$.
In PIM kinematics the pair is treated as a unit so that, on the partonic level,
we have
\begin{eqnarray}
\label{qq_PIM}
i(k_1) + j(k_2) &\longrightarrow& Q \overline Q(p') +
X(k')\, . 
\end{eqnarray}
The square of the heavy quark pair mass is $p'^2 = M^2$.
At partonic threshold,
$X(k') = 0$, the three Mandelstam invariants are
\begin{eqnarray}
\label{tupidef}
s = M^2 \, ,\quad  t_1 = - \frac{M^2}{2} ( 1 - \beta_M\, \cos \theta )
\, ,\quad u_1 = - \frac{M^2}{2} ( 1 + \beta_M\, \cos \theta )\, 
\end{eqnarray}
where $\beta_M=\sqrt{1-4m^2/M^2}$ and $\theta$ is the scattering
angle in the parton center of mass frame.  Now the distance from threshold
is $x = 1 - M^2/s \equiv 1-z$ where $z = 1$ at threshold. In PIM kinematics
the cross sections are functions of $M^2$ and $\cos \theta$.

The resummation is done in moment space by making a Laplace
transformation with respect to $x$, the distance from threshold.  Then the
singular functions become linear combinations of $\ln^k \tilde{N}$ with
$k \leq l+1$ and $\tilde{N} = Ne^{\gamma_E}$ where $\gamma_E$ is the Euler
constant.  The 1PI resummed double differential partonic
cross section in moment space is 
\begin{eqnarray}
\label{resum:eq:9}
&& \hspace{-5mm}
s^2 \frac{d^2 \sigma^{\rm res}_{ij}(\tilde{N})}
{dt_1\,du_1}
\,=\,{\rm Tr}\Bigg\{
H_{ij} 
%\\ \nonumber & &\hspace{0mm} \times
{\rm \bar{P}} \exp\left[\int\limits_{m}^{m/\tilde{N}} {d\mu'\over\mu'} 
(\Gamma^{ij}_S\left(\alpha_s(\mu^{\prime})\right))^{\dagger}\right] \!
{\tilde S}_{ij}%\!\left(1\right)
{\rm P} \exp\left[\int\limits_{m}^{m/\tilde{N}} {d\mu'\over\mu'} 
\Gamma^{ij}_S\left(\alpha_s(\mu^{\prime})\right)\right] \! \Bigg\} \\
\nonumber & & \times
\exp\left(\tilde{E}_{i}(\tilde{N}_u,\mu,\mu_R)\right)\, 
\exp\left(\tilde{E}_{j}(\tilde{N}_t,\mu,\mu_R)\right)\,\, \exp\Bigg\{ 2\,
\int\limits_{\mu_R}^{m}{d\mu'\over\mu'}\,\, \Bigl(
\gamma_i\left(\alpha_s(\mu^{\prime})\right) +
\gamma_j\left(\alpha_s(\mu^{\prime})\right) \Bigr) \Bigg\}\, .
\end{eqnarray}
To find the PIM result, transform $t_1$ and $u_1$ to $M^2$ and $\cos \theta$ 
using
Eq.~(\ref{tupidef}). 
%and multiply the right-hand side by $\beta_M/2$.  The
%trace is in a space which couples SU(3) of the partons to singlets, 2 
%dimensions for $q \overline q$ and 3 for $gg$. 
$(\overline {\rm P})$ P refer to
(anti-)path ordering in scale $\mu'$.  The cross section depends on the `hard',
$H_{ij}$, and `soft', $\tilde{S}_{ij}$, hermitian matrices.  The `hard' part
contains no singular functions.  The `soft' component contains the singular 
functions and, from its renormalization group equation, the soft anomalous
dimension matrix $\Gamma_S^{ij}$, 2 dimensional for $q \overline q$ and 3 for
$gg$, can be derived. 
%along with its eigenvalues $\gamma_i$.  
The universal Sudakov 
factors, the same in 1PI and PIM, are in the exponents $\tilde{E}_i$, expanded
as
\begin{eqnarray}
\label{resum:eq:8}
\exp(\tilde{E}_{i}(\tilde{N}_u,\mu,m))
\simeq 1 + \frac{\alpha_s}{\pi}\left(\sum_{k=0}^2 C^{i,(1)}_{k}
\ln^k(\tilde{N}_u)  \right)
+ \left(\frac{\alpha_s}{\pi}\right)^2\left(
\sum_{k=0}^4 C^{i,(2)}_{k}\ln^k(\tilde{N}_u)
\right)
+\ldots \, \, .
\end{eqnarray}
The coefficients $C^{i,(n)}_k$, as well as the detailed derivation of the 
resummed and finite-order cross sections, can be found in Ref.~\cite{KLMV}.
The momentum space cross sections to NNLO-NNLL are obtained by gathering terms
at ${\cal O}(\alpha_s^3)$ and ${\cal O}(\alpha_s^4)$, inverting the Laplace
transformation and matching the $\tilde{N}$-independent terms in $H_{ij}$ and
$\tilde{S}_{ij}$ to exact ${\cal O}(\alpha_s^3)$ results.

\begin{figure}[htb] 
\setlength{\epsfxsize=0.95\textwidth}
\setlength{\epsfysize=0.25\textheight}
\centerline{\epsffile{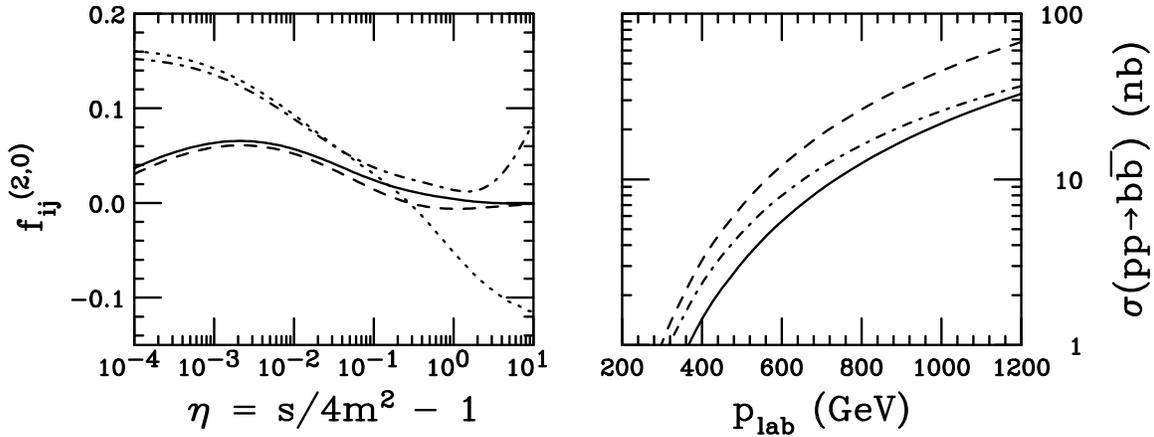}}
\caption[]{The left-hand side shows the $\eta$-dependence 
of the NNLL scaling functions.  We show 
$f^{(2,0)}_{q \overline q}(\eta)$ in 1PI (solid) and PIM (dashed) kinematics
and $f^{(2,0)}_{gg}(\eta)$ in 1PI (dot-dashed)
and PIM (dotted) kinematics. The right-hand side compares the
total $b \overline b$ cross sections at fixed target energies calculated with
CTEQ5M and $\mu  =m = 4.75$ GeV.  The exact NLO result is shown in the 
solid curve while the NNLO-NNLL results for 1PI and PIM kinematics are given
by the dashed and dot-dashed curves respectively.
}
\label{plots} 
\end{figure}

We have studied the partonic and hadronic total cross sections of $t \overline
t$ and $b \overline b$ production.  Any difference in the integrated cross
sections due to kinematics choice arises from the ambiguity of the estimates.
At leading order, no additional soft partons are produced and the threshold
condition is exact.  Therefore, there is no difference between the total cross
sections in the two kinematic schemes.  However, beyond LO and threshold
there is a difference.  To simplify the argument, the total partonic 
cross section may be expressed in terms of dimensionless scaling functions
$f^{(k,l)}_{ij}$ that depend only on $\eta = s/4m^2 - 1$ \cite{KLMV},
\begin{eqnarray}
\label{scalingfunctions}
\sigma_{ij}(s,m^2,\mu^2) = \frac{\alpha^2_s(\mu)}{m^2}
\sum\limits_{k=0}^{\infty} \,\, \left( 4 \pi \alpha_s(\mu) \right)^k
\sum\limits_{l=0}^k \,\, f^{(k,l)}_{ij}(\eta) \,\,
\ln^l\left(\frac{\mu^2}{m^2}\right) \, .
\end{eqnarray} 
We have constructed LL, NLL, and NNLL approximations to $f_{ij}^{(k,l)}$ 
in the $q
\overline q$ and $gg$ channels for $k \leq 2$, $l \leq k$.  Exact results are
known for $k=1$ and can be derived using renormalization group methods for 
$k=2$, $l=1,2$.  Thus the best NNLO estimate of the cross section includes the
exact scaling functions and the NNLL estimate of $f_{ij}^{(2,0)}$.  On the 
left-hand side of Fig.~\ref{plots}, we compare $f_{q \overline q}^{(2,0)}$ and
$f_{gg}^{(2,0)}$, the only approximate scaling function, in 1PI and PIM.  The
results are quite similar at small $\eta$ but begin to differ for $\eta \geq
0.1$, especially in the $gg$ channel.  If the parton flux is maximized for
$\eta < 1$, as for the HERA-B energy, $\sqrt{S} = 41.6$ GeV, the reaction is
close enough to threshold for the results to be reliable.  Unfortunately at
RHIC, the flux peaks at $\eta \approx 1$, making predictions at RHIC from the
threshold approximation alone unreliable.  
The $gg$ channel dominates $b \overline
b$ production in $pp$ interactions.  An inspection of the scaling functions
shows that the results could differ substantially between the two kinematics.

The total hadronic cross section is obtained by convoluting the total partonic
cross section with the parton densities $f_i^p$ evaluated at momentum fraction
$x$ and scale $\mu$,
\begin{eqnarray}
\label{totalhadroncrs}
\sigma_{pp}(S,m^2) = \sum_{i,j = q,{\overline q},g} \,\,
\int_{4m^2/S}^{1}\, d\tau \, \int_\tau^1 \frac{dx_1}{x_1} 
%\int_0^1 dx_2 \delta(x_1 x_2 - \tau) \,
f_i^p(x_1,\mu^2) f_j^p\bigg(\frac{\tau}{x_1},\mu^2\bigg) \, 
\sigma_{ij}(\tau S,m^2,\mu^2)\, .
\end{eqnarray}
If the peak of the convolution of the parton densities is at $\eta < 1$, the
approximation should hold.  On the right-hand side of Fig.~\ref{plots} we 
compare the 1PI and PIM results with the exact NLO results, all calculated
with the CTEQ5M parton densities \cite{cteq5}
and $\mu = m$.  The NNLO-NNLL corrections
are substantial.  The average of
the NNLO-NNLL 1PI and PIM cross sections, 
\begin{eqnarray}
\sigma_{b \overline b}({\rm 41.6 \, GeV}) = 30 \pm 8 \pm 10 \, {\rm nb} \, \, ,
\end{eqnarray} 
is in good agreement with the $b \overline
b$ total cross section measured by HERA-B \cite{herab}.  The first uncertainty
is due to the kinematics choice, the second to the scale dependence.  The
uncertainties in scale and kinematics choice are essentially equivalent.

\end{document}